# Novel Cobalt-Based Magnetic Weyl Semimetals with High-Thermodynamic Stabilities

Wei Luo[1,2], Yuma Nakamura[2], Jinseon Park[2], Mina Yoon[1,2*]

[1]Center for Nanophase Materials Sciences, Oak Ridge National Laboratory, Oak Ridge, TN 37831, U.S.A.

[2]Department of Physics and Astronomy, University of Tennessee, Knoxville, TN 37916, U.S.A.

*Email: *myoon@ornl.gov*

**Abstract**

Recent experiments identified $Co_3Sn_2S_2$ as the first magnetic Weyl semimetal (MWSM) [Science 365, 1282 (2019), Science 365, 1286 (2019)]. Using first-principles calculation with a global optimization approach, we explore the structural stabilities and topological electronic properties of cobalt (Co)-based shandite and alloys, $Co_3MM'X_2$ (M/M'=Ge, Sn, Pb, X=S, Se, Te), and identify new stable structures with new Weyl phases. Using a tight-binding model, for the first time, we reveal that the physical origin of the nodal lines of a Co-based shandite structure is the interlayer coupling between Co atoms in different Kagome layers, while the number of Weyl points and their types are mainly governed by the interaction between Co and the metal atoms, Sn, Ge, and Pb. The $Co_3SnPbS_2$ alloy exhibits two distinguished topological phases, depending on the relative positions of the Sn and Pb atoms: a three-dimensional quantum anomalous Hall metal, and a MWSM phase with anomalous Hall conductivity ($\sim$1290 $\Omega^{-1}cm^{-1}$) that is larger than that of $Co_2Sn_2S_2$. Our work reveals the physical mechanism of the origination of Weyl fermions in Co-based shandite structures and proposes new topological quantum states with high thermal stability.

**Introduction**

Recent years have seen tremendous development in topological quantum materials (TQMs), including topological insulators (TIs) and semimetals with nontrivial band topology. However, there remain a few challenges to practical applications of TQMs. For example, the challenge in the development of TIs,[1-5] one of the most widely studied classes of topological materials, is to protect their key features against disorders or defects that may destroy their quantized conductance.[6, 7] On the other hand, the discovery of new Chern insulators[8-11] with broken time-reversal symmetry is a challenge because of the incompatibility between ferromagnetism and electronic insulation. Topological semimetals (TSMs), defined using a local topological invariant, have become a hot spot in the family of topological materials because of their rich physical properties. One of the most important TSMs is the Weyl semimetal (WSM), whose band dispersion near the Weyl point can be described by the "Weyl equation."[12] Although nonmagnetic WSMs have been thoroughly studied in recent years,[13-18] magnetic WSMs (MWSMs) are still rare, despite their great potential as building blocks for next-generation ultra-fast topological spintronics.

MWSMs exhibit many exotic physical properties, such as quantized anomalous Hall effects,[19] large anomalous Hall conductivity (AHC),[20-22] and domain wall physics.[23] Since the theoretical discovery of pyrochlore $Y_2Ir_2O_7$,[24] the first MWSM, several other potential MWSMs have been reported. They include $HgCr_2Se_4$,[19] Heusler compounds,[25-29] $Fe_3Sn_2$,[30] $Sn_2Nb_2O_7$,[31] spinel compounds[32] and rare earth materials.[33] Recent experiments confirmed the predicted shandite structure of $Co_3Sn_2S_2$ as an MWSM by directly observing its Fermi-arc and linear dispersion bulk bands.[34, 35] Simultaneously, the magnetic Weyl fermion line was confirmed by directly observing the drumhead surface states in the room-temperature magnet $Co_2MnGa$.[36] However, our understanding of their structural stabilities, microscopic physical mechanism of Weyl points, topological phase transition, and novel properties of cobalt (Co)-based shandites and their alloys is still lacking.

In this letter, we investigate the topological properties of $Co_3MM'X_2$ (M/M'=Ge, Sn, Pb and X=S, Se, Te, denoted as Co-MM'-X for simplicity) compounds with high thermodynamic stabilities. Using a global structure search approach based on a particle swarm optimization (PSO) algorithm, we discovered that their structural stabilities as a shandite phase are well described by a structural tolerance factor—defined by the ratio between the atomic radii of the metal and the chalcogen atoms constituting a compound—below which the shandite structure becomes unstable. Shandite becomes a ground state for Co-Sn-S, Co-SnPb-S, Co-Pb-S, and Co-Pb-Se, metastable for Co-Sn-Se and Co-Ge-S, and unstable for Co-Ge-Se and Te-based systems. We determined that Co-SnPb-S alloys can be three-dimensional (3D) quantum anomalous Hall metals (QAHMs) or MWSMs, depending on the relative positions of the Sn and Pb atoms. The MWSM phase had AHC of 1290 $\Omega^{-1}$ cm$^{-1}$ near the Fermi level, even higher than that of $Co_3Sn_2S_2$. We established a tight-binding (TB) model that reproduced the number and types of Weyl points for a Co-M-X system, analyzed by first-principles calculations. Our TB model explains that the formation of nodal lines originates from the interlayer coupling between Co atoms in different Kagome layers, and the number and types of Weyl points are mainly governed by the interaction between interlayer M' and Co atoms.

**Results**

*Crystal structures and magnetic ground states for Co-MM'-X*

Figure. 1a depicts the crystal structure of a shandite compound, Co-MM'-X, in which the cobalt atoms form a Kagome lattice. Figure 1b depicts the atomic configurations projected on the Kagome lattice. Metal atoms occupy both the in-plane sites of the Kagome lattice and the sites in between the Kagome lattice layers; metals on each site are labeled as M and M'. Figure 2a summarizes the stable and metastable crystal phases of Co-MM'-X (M/M'=Ge, Sn, Pb and X=S, Se, Te) found by PSO[37, 38] based on first-principles density functional theory (DFT) calculations (see Methods section). The crystal structures are color coded and arranged by the energy hierarchy for a given composition; i.e., the one listed first is the most stable configuration. Here, the frequency indicates the number of times the given crystal was identified; thus, it relates the probability of discovering each structure. The energy differences (ΔE) between the shandites and the ground state structures for the given compounds are listed in Fig. S1. The compounds Co-Sn-S, Co-Pb-S, Co-Pb-Se, and Co-SnPb-S have the shandite structure (the space group166, red bar in Fig. 2a) as their most stable structure, whereas the shandite structure becomes unstable for Co-Ge-Se and Co-M-Te. For Co-Sn-Se, Co-Ge-S, Co-GeSn-S, and Co-GePb-S, other structures are the most stable ones, with energy differences of less than ~60 meV per formula compared with that of the shandite, showing that the shandite is metastable for those compounds. Note that moderately metastable materials, with energies above the ground state commensurate with the found metastability range (~70 meV/atom, varying by chemistry and composition), are also reasonable candidates for synthesis.[39]

We discovered that the ratio between the sizes of the metal atoms and the sizes of the chalcogens is related to the structural stability of a shandite; and we introduced a structure tolerance factor (*s*), *s*=r(M)/r(X) as a structure descriptor for shandite, where r(M) and r(X) indicate the radius of a metal and a chalcogen, respectively, and the average atomic radius of two metal atoms is taken as r(M) for metal alloys. As demonstrated in Fig. 2a, only the compounds with large *s* values (≥1.5) adopt shandite as a ground state structure. The relationship between the tolerance factor *s* and the stability of the shandite structure is presented in Fig. 2b. We further analyzed the magnetic stabilities categorized in the ferromagnetic (FM), A-type anti-ferromagnetic (AFM), C-type AFM, and nonmagnetic states (see Fig. S2 of SI), and concluded that all the shandite structures maintain an FM ground state with the spin oriented along the z direction (the energy differences between different magnetic states are summarized in Table. S1). Given the structural and magnetic stabilities, we propose Co-Pb-S and Co-SnPb-S as new magnetic shandite compounds that may be synthesized by experiments. Interestingly, recently, the alloy system $Co_3Sn_{2-x}In_xS_2$ has been synthesized[40, 41] by the Bridgeman technique[42] and was identified to maintain a large AHC (1500±300 $\Omega^{-1}cm^{-1}$ at *x*=0.15). Our calculations also confirmed the stability of Co-Sn-S, as it has already been synthesized and identified as an FM Weyl semimetal.[43] In the following discussion, we mainly focus on the Co-MM'-S systems.

*Weyl points in Co-M-S (M=Ge, Sn, Pb) systems*

The shandite Co-Sn-S has nodal lines located at the $k_b=k_c$ ($k_b$ and $k_a$ range from $-\frac{\pi}{2}$ to $\frac{\pi}{2}$) mirror plane ($M_{010}$ in real space) in the Brillouin zone (BZ) [20, 22, 44] (see the planes highlighted in Fig. 1c; those three mirror planes are equivalent owing to the $C_{3z}$ rotation symmetry along the z direction). We first investigated the band structures of compounds Co-M-S (M=Ge,Sn,Pb) that have FM ground states with the easy axis perpendicular to the Kagome plane (z direction). The band structures without spin-orbit coupling (SOC) included confirmed that they are all nodal-line semimetals (see Fig. S3), with nodal lines (for spin-up subspace) formed in the $k_b=k_c$ mirror plane (centered at the L point); and they split into Weyl points as SOC is introduced. Figure 3a and 3b show two-dimensional (2D) band structures with SOC in $k_b=k_c$ mirror plane, where the Weyl points—crossing points between valence bands and conduction bands—are highlighted by the red dots. By plotting the inverses of energy gaps between the conduction and valence bands, we can clearly identify the existence of the Weyl points (Fig. 3c). On the $k_b=k_c$ mirror plane, Co-Ge-S and Co-Sn-S each have a pair of Weyl points, while Co-Pb-S has three pairs. Note that the Weyl points in each pair are symmetric with respect to the inversion center (L point) in the BZ, thus they maintain opposite chirality. Figure 3d presents 1D band structures, tangential to the nodal lines at Weyl points, that identify type-II Weyl points[14] for Co-Ge-S and Co-Pb-S, and type-I Weyl points for Co-Sn-S. Given the $C_{3z}$ symmetry of the system, there are three pairs of type-II Weyl points in the BZ (another two pairs of Weyl points are located at the $k_a=k_b$ and $k_a=k_c$ mirror planes) for Co-Ge-S, nine type-II pairs for Co-Pb-S, and three type-I pairs for Co-Sn-S. Our results for Co-Sn-S are consistent with theory[20, 22, 44] and experiments[34, 35] in the literature. The compound Co-Pb-S had two pairs of Weyl points so close that they almost formed short nodal lines; resolving those Weyl points might be an experimental challenge. We further explored the whole BZ in search of Weyl points located at general points and found additional Weyl points for Co-Pb-S. They were two nonequivalent type-I Weyl points, one located close to the Fermi level (~5 meV higher) and the other one at ~88 meV above the Fermi level. Considering $C_{3z}$ and mirror symmetries, there were 12 pairs of Weyl points located at general points in the BZ. The type, energy level, position, chirality (characterized by Berry curvature around each Weyl points, see Fig. S4) and number of Weyl points Co-M-S systems are summarized in Table S2.

***3D quantum anomalous Hall phase and large AHC in Co-based alloy systems.***
The shandite Co-MM'-S (M/M'=Sn and Pb) has two distinctive configurations, depending on the occupation of Sn and Pb atoms at either at in-plane (M) or interlayer (M') sites of the Kagome plane (see Fig. 1a). Both structures have shandite FM ground states (see Table S1 of SI), and the formation enthalpy of the Co-SnPb-S is lower than that of Co-PbSn-S by ~45 meV per formula unit. Interestingly, these two configurations display very distinctive topological properties: Co-SnPb-S belongs to the 3D quantum anomalous Hall phase,[45, 46] whereas Co-PbSn-S is an MWSM with AHC (~1290 $\Omega^{-1}\text{cm}^{-1}$) higher than that of Co-Sn-S, as explained below in detail.

Figure 4a compares the band structures of Co-SnPb-S and Co-PbSn-S in the $k_b=k_c$ mirror plane. Both systems are nodal-line semimetals if the SOC effects are not considered (see Fig. S3). In the case of Co-SnPb-S, the SOC induces gaps in the nodal line without creating any Weyl points, as indicated in the inverse-gap plot of the $k_b=k_c$ mirror plane of Fig. 4b. The compound

Co-SnPb-S with a finite local band gap becomes more stable than the metallic compound Co-PbSn-S. We further confirmed that no Weyl points existed at other general points by performing dense k-points calculations (401×401×401) employing the Wannier interpolation method (see Methods section).

In a stark contrast to the Co-SnPb-S with no Weyl points, Co-PbSn-S contained type-II Weyl points with SOC. There was a pair of type-II Weyl points (Fig. 4b, $W_1^-$ and $W_1^+$; see Fig.S5a for their types and energy levels) in the $k_b=k_c$ mirror plane and another pair ($W_2^-$ and $W_2^+$) located at the $k_a=k_b=k_c$ axis (i.e., the z axis in real space) in the BZ with opposite chirality (the specific coordinate of this pair of Weyl points can be seen in Table S2). Note that near this pair of Weyl points, we found that a strict flat band pierces two bands (these conduction and valance bands cross each other, leading to the type-II Weyl point in the $k_a=k_b=k_c$ axis [Fig. 4b, $W_2^-$ and $W_2^+$] ) and formed another four critical-tilt Weyl points in the $k_a=k_b=k_c$ axis (Fig. S5b). The physical mechanism underlying the Weyl points formed from 3D flat band has been demonstrated before.[31] This kind of critical-tilt Weyl point may be a good candidate for investigating the strongly correlated Weyl physics. The calculated AHC for Co-PbSn-S was as high as 1290 $\Omega^{-1}cm^{-1}$, higher than that of Co-Sn-S. See the AHC of the Co-based systems in Fig. S6. A recent paper[36] reports $Co_2MnGa$ that has a giant AHC (~1530 $\Omega^{-1}cm^{-1}$) at the energy ~50 meV below the Fermi level. The compound Co-PbSn-S had a peak in AHC closer to the Fermi level (~15 meV above the Fermi level) (Fig. 4c), which makes it a promising candidate for topological spin-transport devices.

On the other hand, Co-SnPb-S has another distinctive topological property—a 3D QAHM; its band structure can adiabatically evolve into a 3D quantum anomalous Hall insulator[45] (QAHI) without any band crossing.[47, 48] To confirm its nontrivial band topology, we followed the criteria proposed by Jin et al.[45]: (1) the first Chern number of a 2D cut with four time-reverse invariant momenta (TRIM) points is a non-zero integer, and (2) the product over the inversion eigenvalues of all occupied bands at eight TRIM points must be 1. We calculated the first Chern number for Co-SnPb-S in a time reversal invariant plane ($k_a$=0 plane with $k_b$ and $k_c$ range from $-\frac{\pi}{2}$ to $\frac{\pi}{2}$) as 1 ($C|_{k_a=0} = 1$); i.e., it is a 2D Chern insulator. Then we evaluated the parities for all the occupied bands at eight TRIM points (see Table S3), which also satisfied the second condition. Thus, we conclude that Co-SnPb-S is a QAHM with an AHC of ~530 $\Omega^{-1}cm^{-1}$, much lower than that of Co-PbSn-S, perhaps because of the lack of Weyl points (Fig. 4b).

**Discussion**

*Tight binding model for the Co-based shandite structure.*

We analyzed the contribution of each atomic species to the emergence of nodal lines through band inversion between the conduction and valence bands. For all the compounds—Co-Sn-S, Co-SnPb-S, and Co-PbSn-S—the $d_{xy}$ and $d_{z^2}$ orbitals of Co atoms were the main contributors to the formation of the nodal lines, with some contributions from the $p_z$ orbitals of metals located at the interlayer sites (See Fig. S7), negligible contributions from any metals located at the intralayer sites and the S atoms for the band inversion. We tried to use $d_{xy}$ and $d_{z^2}$ orbitals for each Co atom to construct the TB model. However, we find it can not reproduce the nodal line in the $k_b=k_c$ mirror plane. We noted that the crystal field which surrounds the Co atom maintains a low symmetry. Thus, the five $d$ orbitals of a Co atom will split and each of them belongs to 1D irreducible representation, with orbital moments quenched ($L_{eff}=0$)[49]. When the onsite SOC is "turned on," the total angular momentum equals to 1/2 ($J_{eff}=1/2$). Hence, we try to use a pseudospin $J_{eff}=1/2$ ($J_{1/2}$) orbital[50, 51] to construct our TB model.

Our TB Hamiltonian included interactions between $J_{1/2}$ orbitals on each Co atom, $p_z$ orbitals of interlayer metal atoms, and $p_z$ orbitals of S atom. For simplicity, we assumed an infinitely large exchange field and limited our discussion in the spin-majority subspace, which reduced to six orbitals in total in the unit cell. Our TB Hamiltonian reads

$$H = H_{Co} + H_{soc} + H_{Rashba} + H_{Co-M} + H_{Co-S} + +H_{onsite}. \tag{1}$$

The first term, $H_{Co}$, describes the hopping interactions between the $J_{1/2}$ orbitals of Co atoms:

$$H_{Co} = \sum_{<ij>\alpha} t_1 c_{i\alpha}^\dagger c_{j\alpha} + \sum_{\ll ij \gg \alpha} t_2 c_{i\alpha}^\dagger c_{j\alpha} + \sum_{<ij>\alpha} t_3 c_{i\alpha}^\dagger c_{j\alpha}, \tag{2}$$

where $c_{i\alpha}^\dagger$ and $c_{j\alpha}$ represent creation and annihilation of an electron on the $\alpha$ ($J_{1/2}$) orbital at $i$ site and $j$ site, respectively; $t_1$ and $t_2$ are the nearest neighbor (NN) and next nearest neighbor (NNN) hopping parameters between the orbitals in the Kagome plane (see Fig. 1), respectively; $t_3$ is the NN hopping parameter between the interlayer orbitals (see Fig. 1); and $<ij>$ and $\ll ij \gg$ represent the NN and NNN sites for each, respectively. The effective SOC interaction between the Co atoms is incorporated in the second term, $H_{soc}$, as the interaction between $J_{1/2}$ orbitals in the Kagome plane:

$$H_{soc} = it_{so} \sum_{<ij>\alpha} v_{ij} c_{i\alpha}^\dagger c_{j\alpha}, \tag{3}$$

where $t_{so}$ is the effective SOC hopping strength between the NN J$_{1/2}$ orbitals and the sign ($v_{ij}=\pm 1$) depends on the position of an intralayer M atom relative to the Co-Co bond[52] (note that the NNN SOC interaction is negligible). In stark contrast to the hexagonal systems, such as silicene,[52] in which the effective NN SOC interaction vanishes and the next nonzero interaction term is from the NNN interactions, the effective NN SOC of the Co-based system interaction was preserved by the intralayer M atoms that formed a triangle sublattice embedded in the Kagome lattice of Co atoms.

The third term, $H_{Rashba}$, is the NN Rashba SOC term expressed as

$$H_{Rashba} = it_R \sum_{<ij>} u_{ij} c_{i\alpha}^\dagger (\vec{\sigma} \times \vec{d_{ij}^0})^z c_{j\alpha}, \tag{4}$$

where $t_R$ is an effective Rashba hopping strength between the NN J$_{1/2}$ orbitals of Co atoms; $(\vec{\sigma} \times \vec{d_{ij}^0})^z$ represents the z component of a vector product between the Pauli matrix and unit vector of Co-Co bond, $d_{ij}^0$; and the sign of the term ($u_{ij} = \pm 1$) depends on the position of the interlayer M atom relative to the NN Co atoms. $H_{Co-M}$ is the interaction between the nearest Co and interlayer M atom orbitals, i.e., J$_{1/2}$ and $p_z$ orbitals, respectively:

$$H_{Co-M} = \sum_{<ij>\alpha\beta} t_4 c_{i\alpha}^\dagger c_{j\beta}, \tag{5}$$

where $\alpha$ and $\beta$ represent the J$_{1/2}$ orbital of Co and the $p_z$ orbital of the M atom at the interlayer site, respectively, and $t_4$ represents the hopping parameter between the nearest orbitals. The term, $H_{Co-S}$, describes the interaction between Co and S atoms:

$$H_{Co-S} = \sum_{<ij>\alpha\gamma} t_5 c_{i\alpha}^\dagger c_{j\gamma}, \tag{6}$$

where $\gamma$ is the $p_z$ orbital of the S atom and $t_5$ represents the hopping parameter between the nearest orbitals. The term $H_{onsite}$ represents the onsite energies of $p_z$ orbitals of M and S atoms:

$$H_{onsite} = \varepsilon_M \sum_i c_{i\beta}^\dagger c_{i\beta} + \varepsilon_S \sum_i c_{i\gamma}^\dagger c_{i\gamma},$$

where $\varepsilon_M$ and $\varepsilon_S$ are the onsite energies of the $p_z$ of M and $p_z$ of S atoms, respectively. Here, we set the onsite energies of J$_{1/2}$ of Co to be zero.

Next, we investigated the role of each term of the Hamiltonian in the emergence of nodal lines, as well as the formation of Weyl points. The first two terms of $H_{Co}$ in Eq. (2) describe the interactions between Co atoms in the Kagome lattice plane. We failed to capture nodal lines with those terms solely (see Fig. S8a), but with the addition of the interlayer interaction (the third term in Eq. [2]), we were able to monitor the emergence of the nodal lines centered at the L point of BZ (Fig. S8), even without considering any additional terms from metal atoms. Upon application of the NN SOC interaction between the Co atoms in the Kagome plane (Eq. [3]), the nodal line split into two pairs of Weyl points in the $k_b=k_c$ mirror plane. Those Weyl points were robust against further application of the Rashba SOC (Eq. [4]). We found it interesting that the interaction between metals and Co atoms (Eq. [5]) resulted in the correct number of Weyl points, as well as their types, as identified by DFT (Fig. 5a and b): one pair for Co-Ge-S and Co-Sn-S, three pairs for Co-Pb-S, and their types (type-I and type-II). Finally, the last term in the Hamiltonian (Eq. [6]), the interaction between S and Co, slightly modified the shape of the nodal lines and the types of Weyl points in the mirror plane. In summary, the nearest interlayer coupling between $J_{1/2}$ effective orbitals of Co atoms resulted in the emergence of the nodal lines, whereas the number of Weyl points and their types were mainly determined by the interaction between $p_z$ orbitals of M and $J_{1/2}$ of Co. Additionally, the number of Weyl points and their types were sensitive to the on-site energy of $p_z$ orbitals ($\varepsilon_M$) of the M atoms. Figure 5c presents the phase diagram of a Co-based shandite structure in the parameter space of $t_4$ and $\varepsilon_M$, where the number and type of Weyl points are those located on the $k_a=k_b$ mirror plane. At the boundary between W-I and W-II regions, there is a phase transition from type-I to type-II, in which the band crossing is close to a critical tilt. Our DFT results revealed that Co-Sn-Se holds this kind of critical tilt band dispersion at this boundary (see Fig. S5c). These results indicate that our TB captured the essential physical mechanisms that governs the emergence of nodal lines as well as Weyl points in CoMS (M=Ge, Sn, Pb) systems. It is noted that although our TB model does consider the role of intralayer metal site, we can describe the QAHM state which the alloy system Co-SnPb-S maintains. However, for Co-PbSn-S alloy, based on our TB model, with changing the parameters ($t_4$ and $\varepsilon_M$), we did not find the pair of Weyl points which locate at the z axis ($W_2^-$ and $W_2^+$, see Fig. 4(b)). We guess this may be attributed to two reasons. One is that Weyl points emerge occasionally with fine-tuned parameters. Hence, it is challengeable to find them if the proper parameters are not given. The phase diagram (Fig. 5(c)) is calculated with fixing some parameters ($t_1 = -1$, $t_2 = -0.4$, $t_3 = 1.3$, $t_{so} = -0.1$, $t_{rashba} = -0.1$, $t_5 = -1.0$, and $\varepsilon_S = -4.0$), and only $t_4$ and $\varepsilon_M$ are allowed to change. These fix parameters may also limit the finding of the pair of Weyl point at the z axis. Another reason may be from the TB model itself. The Co-based WSMs are novel materials with complex band structure whose bands are highly hybridized with each other. Nevertheless, our tight-binding model is a simplified model without considering the role of intralayer metal site. For the alloy system Co-PbSn-S, the Pb and Sn atoms locate at the intralayer and interlayer site, respectively. Our model does not consider the interaction between Pb (intralayer) and other atoms. This simplification may also limit the finding of the pair of Weyl points ($W_2^-$ and $W_2^+$ in Fig. 4(b)) for Co-PbSn-S system.

In summary, we systematically investigated the stability and electronic properties of shandite Co-based materials. For structural stability, we found a good "descriptor" (i.e., r[M]/r]X]) that can identify not only the stability of the shandite crystal structure but also the stability of the magnetic states. For electronic properties, we found that the behavior of Weyl points was strongly dependent on chemical composition. In particular, the Weyl semimetal Co-PbSn-S exhibited a large AHC very close to the Fermi level and is a good candidate as a spintronic transport device. Finally, with an effective TB model, we revealed that the band inversion nodal line in the Co-based Kagome lattice was derived from the interlayer coupling between Co atoms and that the behavior of Weyl points was mainly determined by the interaction between Co atoms and the interlayer M atoms. Our work gives a clear physical picture of the origination of the Weyl fermion in the Co-based Kagome lattice and provides guidance for synthesizing new stable TQMs.

## Methods

### DFT calculations

Electronic structures were determined by DFT calculations as implemented in the Vienna ab initio simulation package.[53] The projector augmented wave[54, 55] was used to treat the ion-electron interactions, and a generalized gradient approximation of the Perdew-Burke-Ernzerhof[56] functional was applied to consider the exchange-correlation potential. The cutoff energy and k-mesh were 500 eV and 8×8×8, respectively. All the structures were optimized using a conjugate gradient approach with the atomic forces smaller than 0.01 eV/Å.

### PSO simulations

Crystal structure predictions for each chemical composition were carried out by employing the PSO algorithm implemented in the software Crystal Structure Analysis by Particle Swarm Optimization (CALYPSO).[38] For a single prediction, each generation consisted of 30 structures, and structural evolution proceeded to at least the tenth generation. For each generation, 60 percent of the structures were constructed through the PSO algorithm, and the remaining 40 percent were randomly produced to prevent premature convergence of structural prediction. During the structure search process, structural relaxations and total energy calculations were performed using first-principles calculations.

### Anomalous Hall conductivity

First, for $T_3MM'X_2$, we use $3d$ orbitals of T, $5p$ orbitals of M(M'), and $3p$ orbitals of X to construct the TB Hamiltonian as implemented in the Wannier90 package.[57] Then, based on this TB Hamiltonian, we used the WannierTools[58] to calculate the AHC for the $T_3MM'X_2$ system. The AHC can be obtained as the sum of Berry curvatures for the occupied bands[59]:

$$\sigma_{xy} = -\frac{2\pi e^2}{h} \int_{BZ} \frac{d^3(\boldsymbol{k})}{(2\pi)^3} \sum_n f_n(\boldsymbol{k}) \Omega_n^z(\boldsymbol{k}) .$$

Here, $\Omega$ is the Berry curvatures, $n$ is the index of the occupied bands, and $f$ is the Fermi-Dirac distribution function. We used a $150 \times 150 \times 150$ k-points grid to calculate the value of AHC.

**Effective TB model**

The numerical TB model was implemented in the Pythtb[60] package, which uses the orthogonal TB approach. The Bloch-like basis function can be constructed as follows:

$$|\chi_i^k\rangle = \sum_R e^{i\mathbf{k}\cdot(\mathbf{R}+\mathbf{t}_i)}|\phi_{Ri}\rangle.$$

$\mathbf{R}$ and $i$ are the site and orbital index, respectively; $t_i$ is the position of orbital $i$ in the home unit cell; $\phi_{Ri}$ is the TB basis orbital in cell $\mathbf{R}$ and should satisfy the following condition:

$$\langle\phi_{Ri}|\phi_{Rj}\rangle = \delta_{RR'}\delta_{ij},$$

which is the so-called orthogonal TB approach. The eigenstates can be expanded as

$$|\psi_{nk}\rangle = \sum_i C_i^{nk}|\chi_i^k\rangle.$$

Then, the Hamiltonian at $\mathbf{k}$ point can be written as

$$H_{ij}^k = \langle\chi_i^k|H|\chi_j^k\rangle = \sum_R e^{i\mathbf{k}\cdot(\mathbf{R}+\mathbf{t}_j-\mathbf{t}_i)}H_{ij}(\mathbf{R}).$$

Here, the phase factor $e^{i\mathbf{k}\cdot(\mathbf{R}+\mathbf{t}_j-\mathbf{t}_i)}$ is included, which is called convention I.[60] By solving the matrix equation

$$H_k \cdot C_{nk} = E_{nk}C_{nk},$$

we can get the eigenvalue and eigenstate at each $\mathbf{k}$ point and thus get the whole bands in the BZ.

**Data availability**

The data that support the findings of this study are available from the corresponding authors upon reasonable request.

**Acknowledgement**

Research was performed at the Center for Nanophase Materials Sciences, which is a DOE Office of Science User Facility, supported by the US Department of Energy, Office of Science, Basic Energy Sciences (BES), Materials Sciences and Engineering Division, and by the Creative Materials Discovery Program through the National Research Foundation of Korea (NRF) funded by the Ministry of Science, ICT and Future Planning (NRF-2016M3D1A1919181). Computing resources were provided by the National Energy Research Scientific Computing Center, which is supported by the Office of Science of the US Department of Energy under Contract No. DE-AC02-05CH11231.

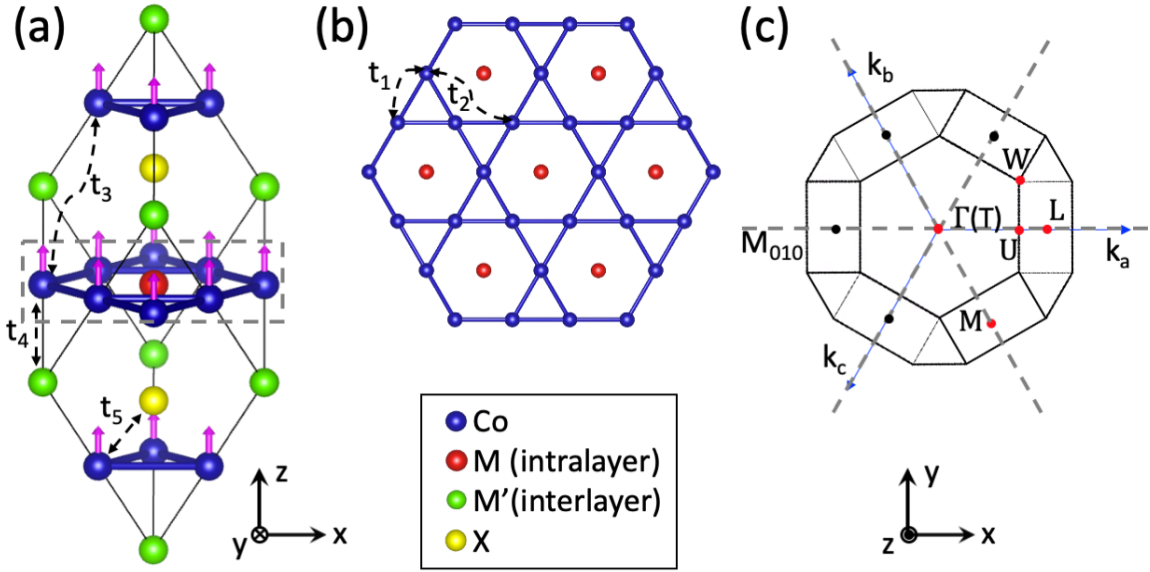

**Figure 1.** (**a**) The crystal structure of shandite Co-MM'-X (the pink arrows represent the orientation of spin). (**b**) The Kagome plane embedded in the shandite structure (highlighted by the dotted grey line in [a]); $t_1$ and $t_2$ represent the NN and NNN hopping parameters between Co orbitals in the Kagome plane, respectively; and $t_3$, $t_4$, and $t_5$ are the NN hopping parameters between the interlayer orbitals of Co atoms, between the orbitals of Co and M', and between the orbitals of Co and X, respectively (see the main text for details). (**c**) The BZ for the shandite structure, where the dotted lines indicate the $k_b=k_c$ mirror plane ($M_{010}$ in real space) as well as the equivalent mirror planes owing to the $C_{3z}$ symmetry.

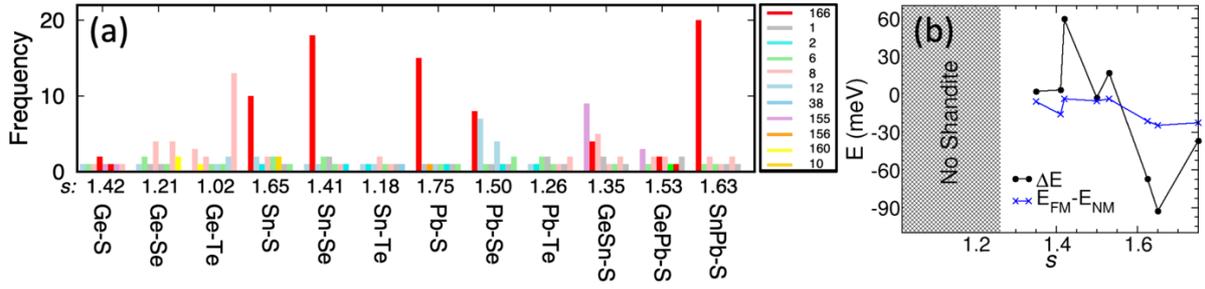

**Figure 2.** (a) The low-energy structures for Co-based materials (Co-M-X or Co-MM'-X), where the crystal structures are color coded and arranged by the energy hierarchy for a given composition, i.e., the one listed first is the most stable configuration. The red bars represent the shandite structure (No. 166). The vertical axis represents the emergent frequency of different structures from the PSO simulation. We introduced a structure tolerance factor ($s$), $s=r(M)/r(X)$ as a structure descriptor for shandite, where r(M) and r(X) indicate the radius of a metal and a chalcogen, respectively, and the average atomic radius of two metal atoms is taken as r(M) for metal alloys. The s values of each compounds are listed. (b) The energy differences ($\Delta E$) between the shandites and the ground state structures and the energy difference between FM and nonmagnetic states ($E_{FM}-E_{NM}$) for the given compounds are plotted in terms of $s$ in black and blue lines, respectively. If the shandite structure is the ground-state structure, $\Delta E$ and $E_{FM}-E_{NM}$ calculate the energy difference to the second stable structure.

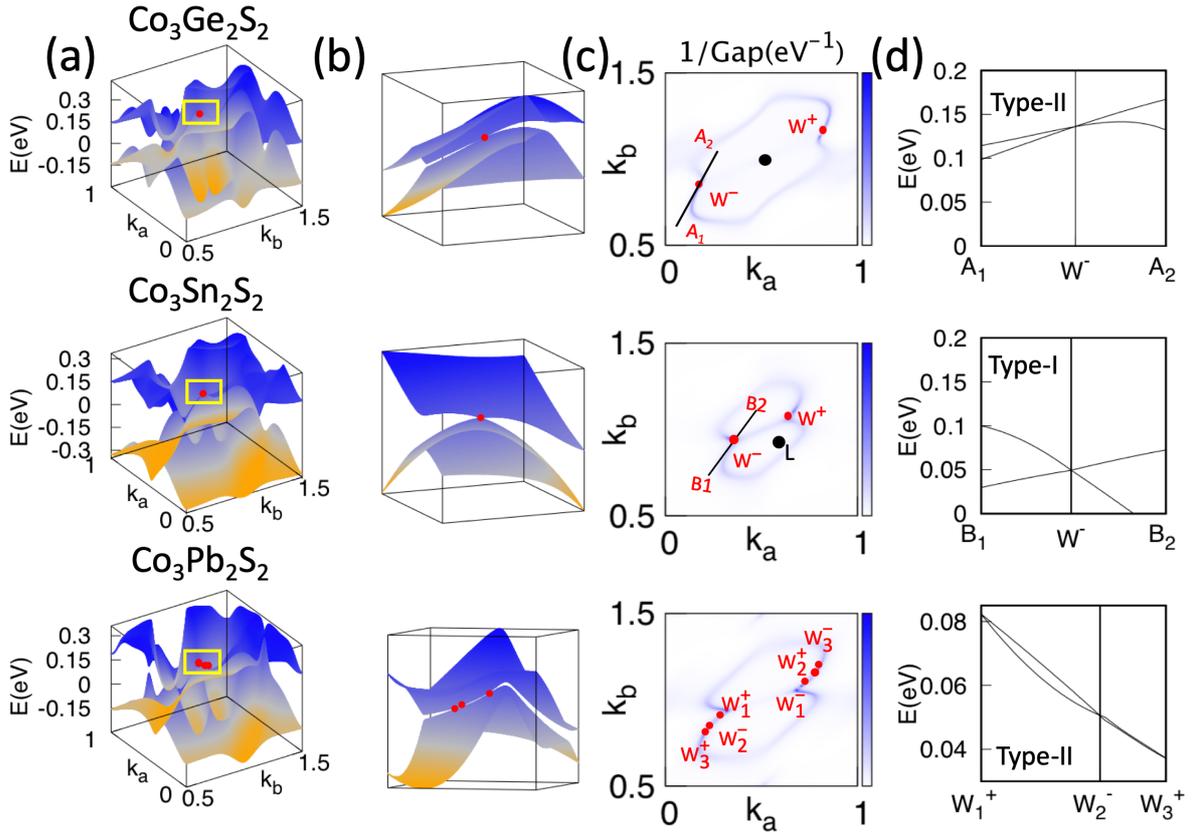

**Figure 3.** (a) 2D plots of conduction and valence bands in $k_b=k_c$ mirror plane for Co-M-S systems. (b) Enlarged data points of the region highlighted by the yellow lines in (a). (c) 2D maps of the inverse energy band gaps (1/gap) in the $k_b=k_c$ mirror plane. If the gap is large, the value of 1/gap is small (white color). Nodal lines open up small bandgaps with SOC, resulting in the high-intensity blue lines in the 1/gap curve. The red points highlight the Weyl points (W) with "+" and "−" signs indicating their chiralities (d) Energy plots along the lines tangential to the Weyl points highlighted in (c), identifying the types of Weyl points. The top, middle, and bottom panels are data for Co-Ge-S, Co-Sn-S, and Co-Pb-S, respectively.

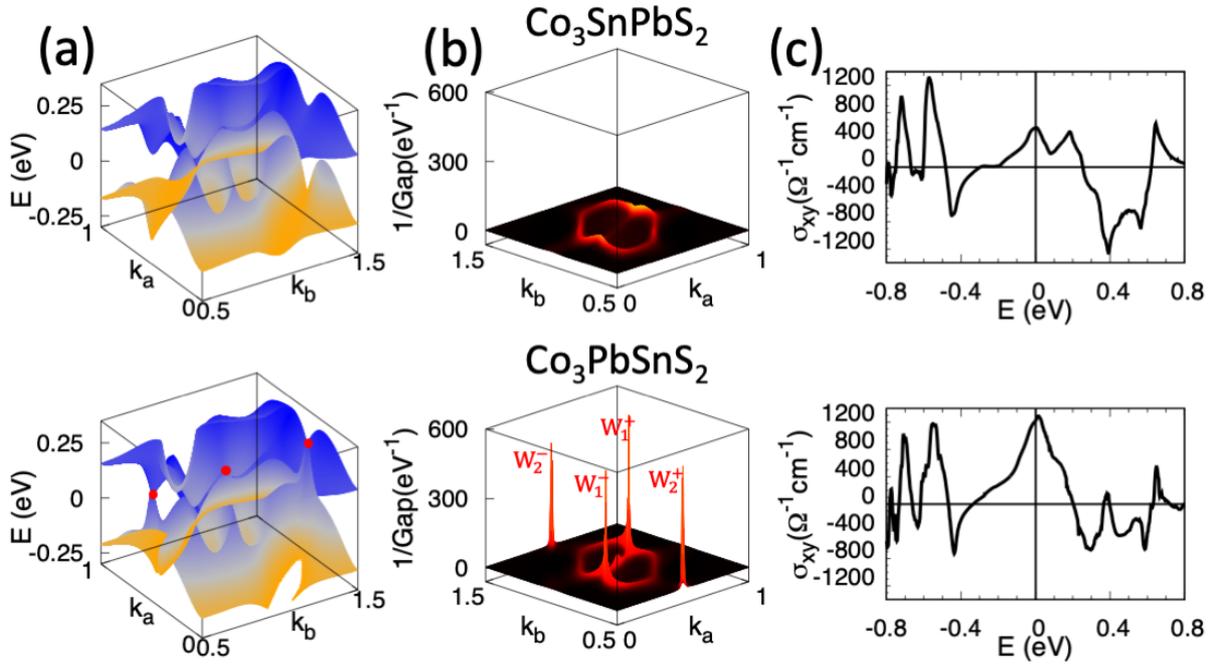

**Figure 4.** 2D band structures, inverse gaps, and anomalous Hall conductivity (AHC) of compounds Co-SnPb-S and Co-PbSn-S on the top and bottom panels, respectively. (a) 2D plots for conduction and valence bands in $k_b=k_c$ mirror plane for systems. Red dots highlight the Weyl points of Co-PbSn-S; there are no Weyl points for Co-SnPb-S. (b) The 3D plot for the inverse of the band gaps in the $k_b=k_c$ mirror plane. Unlike the case of Co-PbSn-S showing Weyl points ($W_1$ and $W_2$), Co-SnPb-S presents energy gaps along the nodal line without leaving any Weyl points in the mirror plane. (c) AHC.

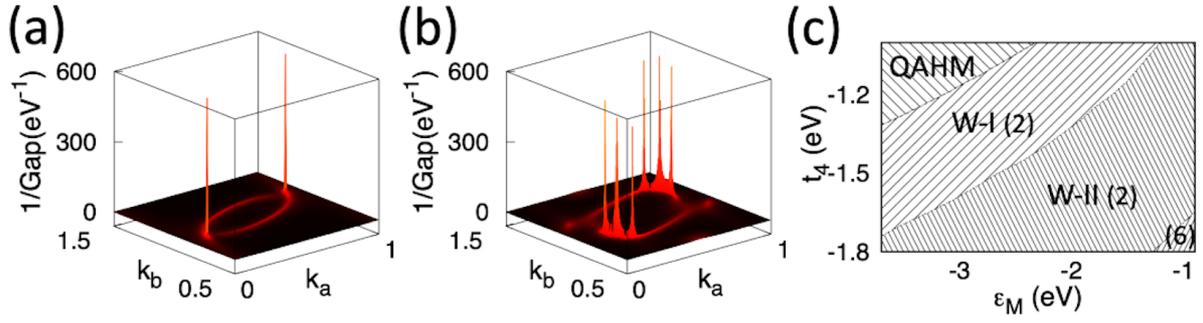

**Figure 5.** (a) A pair of Weyl points in the $k_b=k_c$ mirror plane with parameters $t_1 = -1$, $t_2 = -0.4$, $t_3 = 1.3$, $t_{so} = -0.1$, $t_{rashba} = -0.1$, $t_4 = -1.3$, $t_5 = -0.85$, and $\varepsilon_M = -2.4$. (b) Three pairs of Weyl points in the mirror plane with parameter $t_1 = -1$, $t_2 = -0.4$, $t_3 = 1.3$, $t_{so} = -0.1$, $t_{rashba} = -0.1$, $t_4 = -1.8$, $t_5 = -1.0$, and $\varepsilon_M = -1.0$ (all units in eV). (c) The phase diagram in the $t_4$ and $\varepsilon_M$ parameter space with $t_1 = -1$, $t_2 = -0.4$, $t_3 = 1.3$, $t_{so} = -0.1$, $t_{rashba} = -0.1$, $t_5 = -1.0$, and $\varepsilon_S = -4.0$, where QAHM, W-I, and W-II stand for compounds with quantum anomalous Hall metal, type I, and II Weyl semimetals, respectively; and the numbers in parentheses indicate the number of Weyl points in the $k_b=k_c$ mirror plane. Here all the units of the parameters are in eV.